\documentclass[conference]{IEEEtran}
\IEEEoverridecommandlockouts
\usepackage{cite}
\usepackage{amsmath,amssymb,amsfonts}
\usepackage{algorithmic}
\usepackage{graphicx}
\usepackage{textcomp}
\usepackage{xcolor}
\usepackage{gensymb}
\usepackage{multirow}

\usepackage{caption}% http://ctan.org/pkg/caption
\def\BibTeX{{\rm B\kern-.05em{\sc i\kern-.025em b}\kern-.08em
    T\kern-.1667em\lower.7ex\hbox{E}\kern-.125emX}}
    
\begin{document}

\title{Gradient Backpropagation based Feature Attribution to Enable Explainable-AI on the Edge

\thanks{This work was supported by Semiconductor Research Corporation (SRC) Task 2969.001.\\
To appear in 30th IFIP/IEEE International Conference on Very Large Scale Integration (VLSI-SoC 2022) \hfill}
}

% \vspace{3 cm}

\author{\IEEEauthorblockN{Ashwin Bhat, Adou Sangbone Assoa, Arijit Raychowdhury}
\IEEEauthorblockA{\textit{School of Electrical and Computer Engineering} \\
\textit{Georgia Institute of Technology}, Atlanta, GA, USA \\
(ashwinbhat, aassoa3)@gatech.edu, arijit.raychowdhury@ece.gatech.edu}
}

% \IEEEoverridecommandlockouts
% \IEEEpubid{\makebox[\columnwidth]{978-1-6654-9005-4/22/\$31.00~\copyright2022 IEEE \hfill} \hspace{\columnsep}\makebox[\columnwidth]{ }}
\maketitle

% \IEEEpubidadjcol

\begin{abstract}
There has been a recent surge in the field of Explainable AI (XAI) which tackles the problem of providing insights into the behavior of black-box machine learning models. Within this field, 
\textit{feature attribution} encompasses methods which assign relevance scores to input features and visualize them as a heatmap. Designing flexible accelerators for multiple such algorithms is challenging since the hardware mapping of these algorithms has not been studied yet. In this work, we first analyze the dataflow of gradient backpropagation based feature attribution algorithms to determine the resource overhead required over inference. The gradient computation is optimized to minimize the memory overhead. Second, we develop a High-Level Synthesis (HLS) based configurable FPGA design that is targeted for edge devices and supports three feature attribution algorithms. Tile based computation is employed to maximally use on-chip resources while adhering to the resource constraints. Representative CNNs are trained on CIFAR-10 dataset and implemented on multiple Xilinx FPGAs using 16-bit fixed-point precision demonstrating flexibility of our library. Finally, through efficient reuse of allocated hardware resources, our design methodology demonstrates a pathway to repurpose inference accelerators to support feature attribution with minimal overhead, thereby enabling real-time XAI on the edge.  
\end{abstract}


\section{Introduction}
\begin{figure}[ht!]
\centering
\includegraphics[width=\linewidth]{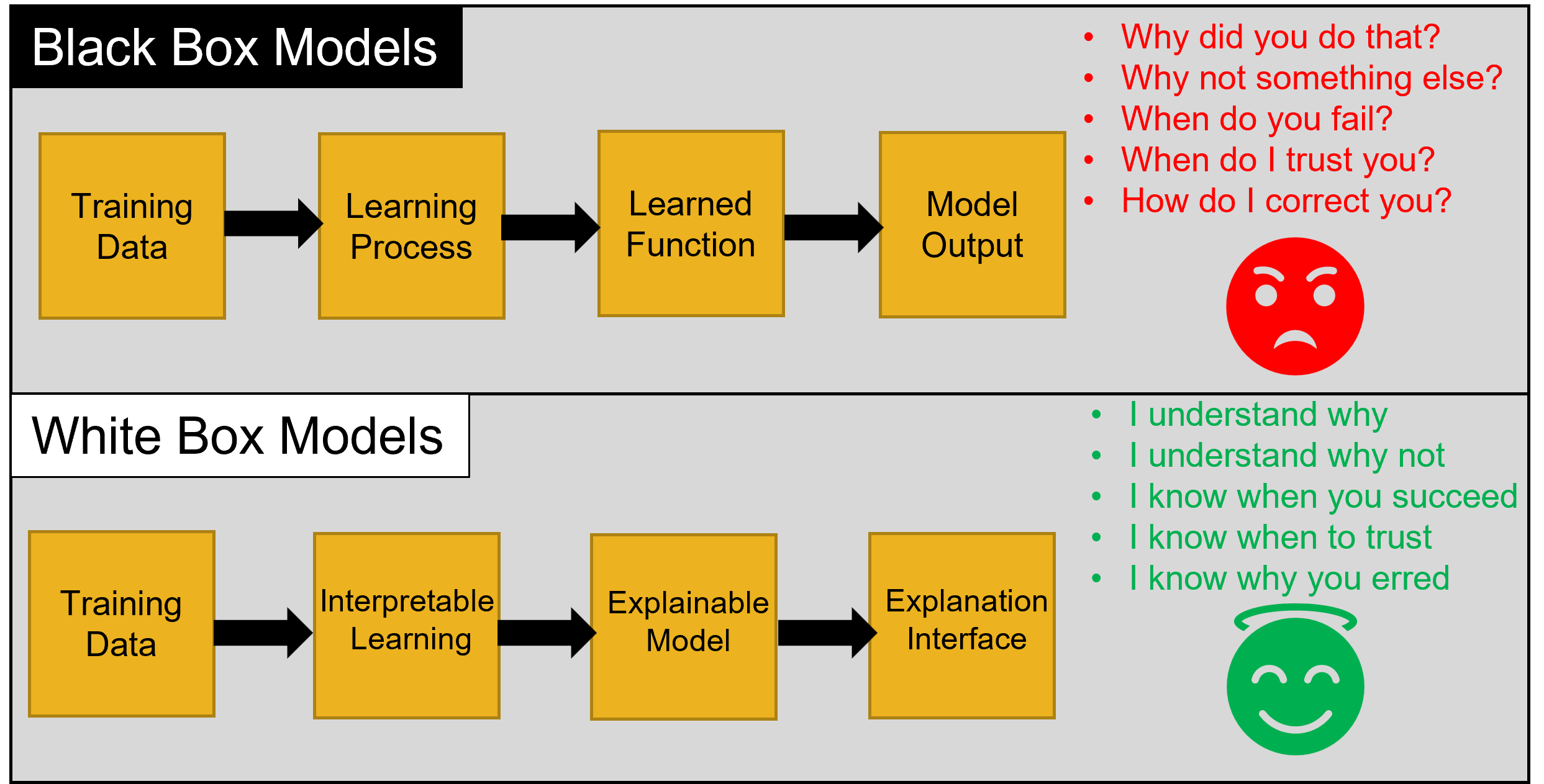}
\caption{Pitfalls of using models as black-box functions v/s advantages of developing Explainable AI.}\label{figure:XAI}
\end{figure}

There has been an exponential surge in the field of machine learning (ML) and artificial intelligence (AI) in the past decade. ML techniques, especially Deep Neural Networks (DNN) have found widespread adoption in various domains such as computer vision, speech recognition, autonomous driving and bio-medical applications. However, one major hurdle currently is the inability to interpret the output of these models since they are treated as a "black-box". The lack of transparency in the model's decision making process severely limits its applicability (Fig \ref{figure:XAI}). In order to address this issue, several techniques have been proposed recently to interpret these models\cite{Adadi2018}. Explainable-AI (XAI) methods shed light into the workings of "black-box" models and thereby identify failure modes, establish trust in the end user and would eventually enable machine teaching\cite{selvaraju2017grad}.

XAI techniques can be broadly classified into three categories namely (1) visualization (2) model distillation and (3) training interpretable models\cite{ras2022explainable}. Among these three, visualization is the only post-hoc explanation method that can be directly applied on existing pre-trained models\cite{ancona2018towards}, and hence is the focus of this work. Visualization comprises of assigning relevance scores to the input features of the model in order to quantify their importance to the output of the black-box model. In the case of image classification using Convolutional Neural Networks (CNN), the feature attribution scores can be visualized as a heatmap of the input pixels. This would highlight regions that contributed most for that particular input-output mapping produced by the model. 

\begin{figure}[ht!]
\centering
\includegraphics[width=\linewidth]{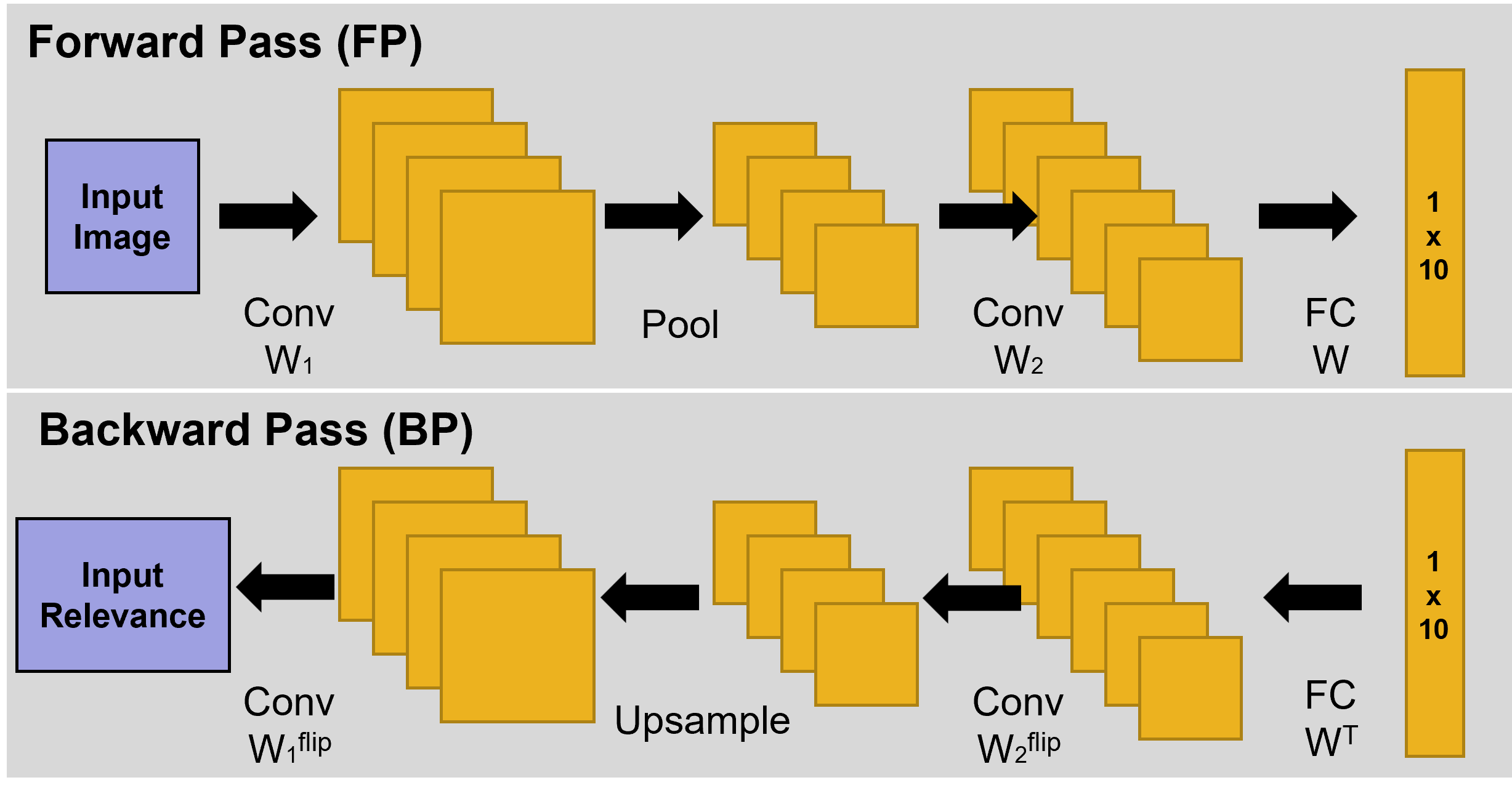}
\caption{The two phases of feature attribution algorithms for CNNs. First, a forward pass to determine inference output. Second, a backward pass to compute activation gradients. }\label{figure:XAIPhases}
\end{figure}

The different feature attribution methods comprise of two common steps. First, a forward pass (FP) through the model to determine the inference result. The second step is a backpropagation (BP) through the model to evaluate the relevance scores for input features and generate the heatmap (Fig. \ref{figure:XAIPhases}). Compared to neural network training, feature attribution does not require calculating gradient with respect to the model parameters for the weight update (WU) step. Thus, the dataflow of feature attribution algorithms (FP+BP) lies in between that of inference (FP) and training (FP+BP+WU).

While inference accelerators have been designed for edge applications, supporting on-device training is challenging because of the large compute and memory overheads. WU is the most expensive step in training. It requires storing all intermediate activations during FP (memory overhead) and calculating gradient with respect to each model parameter (compute overhead). However, feature attribution only involves calculating local activation gradients layer by layer (BP). 

XAI has been deployed for applications such as hardware security\cite{golder2022exploration}, medicine\cite{ibrahim2020explainable} and finance\cite{bhatt2020explainable}. However, challenges remain that prohibit end-users from accessing explanations in real time\cite{bhatt2020explainable}. In this work, we try to answer the question whether real-time XAI can be supported on edge devices. Specifically, by studying the hardware mapping of gradient backpropagation based feature attribution methods, the paper makes the following key contributions:
\begin{itemize}
    \item Dataflow analysis of three gradient backpropagation based feature attribution methods: (1) Saliency Map, (2) DeconvNet, and (3) Guided Backpropagation to determine their h/w resource overhead compared to inference.
    \item We propose a hardware design that efficiently reuses compute blocks and on-chip buffers (designed for inference) during the BP step for feature attribution. The design can be configured to support any of the three feature attribution methods.
    \item We prototype our proposed design on a tiny, resource-constrained FPGA using High-Level Synthesis (HLS), thereby enabling real-time XAI on edge devices.  
\end{itemize}

\section{Feature Attribution}
\begin{figure}[ht!]
\centering
\includegraphics[width=\linewidth]{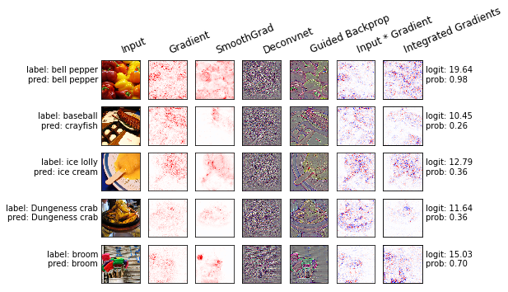}
\caption{An illustration \cite{alber2019innvestigate} of post-hoc feature attribution methods. The generated relevance score are visualized as heatmaps. These heatmaps are visually validated to be highlighting those pixels that are relevant to the model's output decision. In this work, we focus on gradient, deconvnet and guided backprop.}\label{figure:heatmaps}
\end{figure}

\begin{figure}[ht!]
\centering
\includegraphics[width=\linewidth]{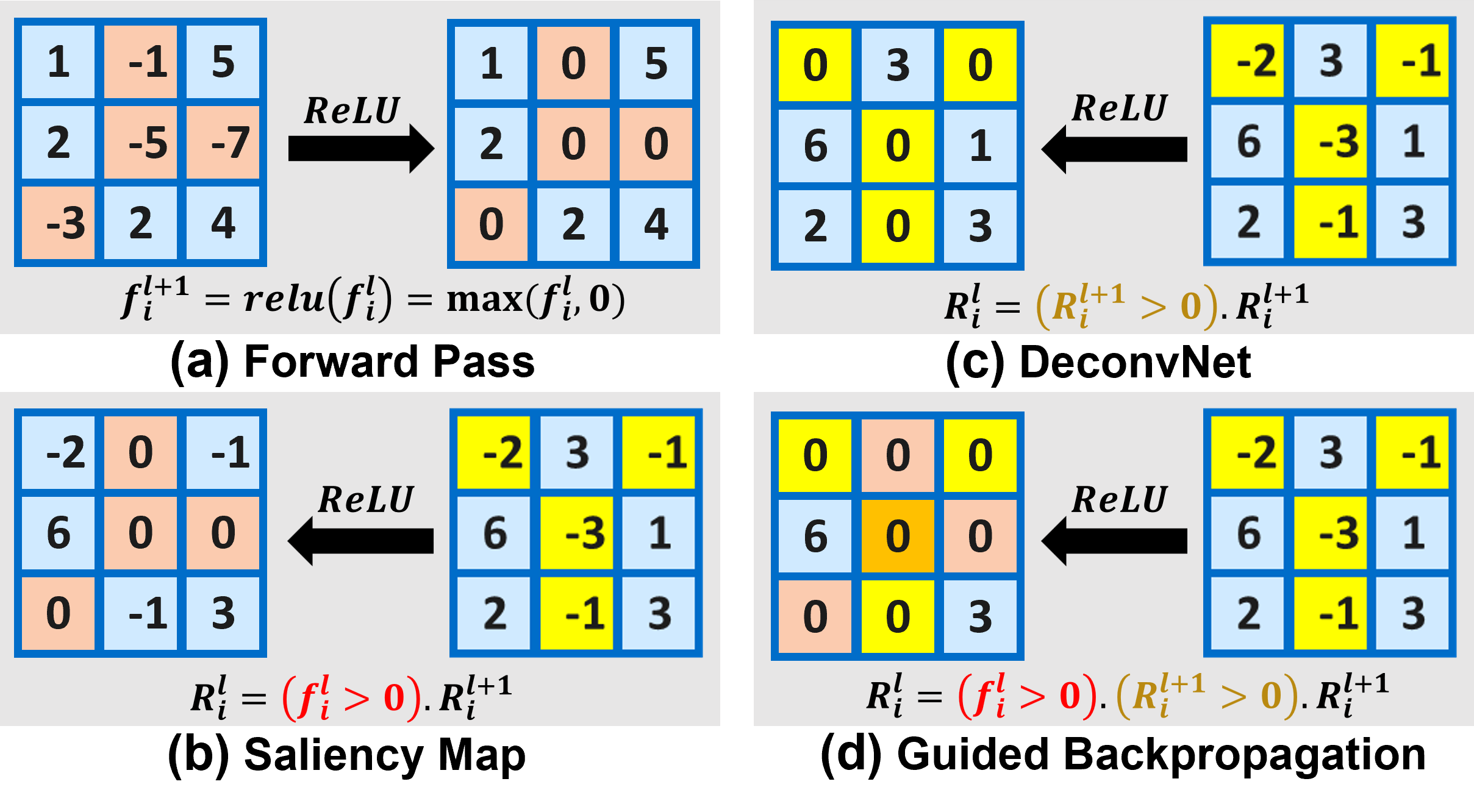}
\caption{Comparison of the dataflow of feature attribution methods at ReLU activation layer (a) Forward pass through a ReLU activation (b) Saliency map performs vanilla gradient computation across ReLU using indices of negative values during forward pass (c) DeconvNet applies ReLU on gradient values (d) Guided Backpropagation combining (b) and (c)  }\label{figure:dataflow}
\end{figure}

Feature attribution methods visualize the contribution of input features to the model's output decision in the form of a heatmap (Fig. \ref{figure:heatmaps}). Higher relevance scores imply that those corresponding features create maximum response or stimulation influencing the model's output. These post-hoc methods can be applied to any off-the-shelf DNN model. After the inference step (FP) to evaluate the output, a backpropagation step (BP) is applied to evaluate gradient signals and pass them from output to input in a layer by layer fashion. We study the dataflow of three different commonly used feature attribution methods: (1) Saliency Map\cite{simonyan2014deep} (2) DeconvNet\cite{zeiler2014visualizing} and (3) Guided Backpropagation\cite{springenberg2014striving}. These methods differ in the handling of the gradient signals when it encounters a ReLU activation (Fig. \ref{figure:dataflow}) layer in the DNN. Equation \ref{eq:ReLU} describes how the network activations are computed during FP when it encounters a ReLU activation at layer L. 

\begin{equation} \label{eq:ReLU}
    f_i^{L+1} = \mbox{ReLU}(f_i^L) = \mbox{max}(f_i^L,0)
\end{equation}

\subsection{Saliency Map}
Saliency map is a baseline gradient based approach which assigns relevance scores (\textit{R\textsubscript{i}(x)}) to input features based on the partial derivative of the model's output (\textit{f\textsubscript{c}(x)}, where \textit{c} is the output class) with respect to each input feature (\textit{x\textsubscript{i}}) as shown in Equation \ref{eq:Saliency}. A large value of the gradient implies that small changes in the value of that input feature would produce large change in the model's output, thereby indicating higher sensitivity. If we consider absolute value of the gradients, the positive and negative contributing features cannot be differentiated.

\begin{equation} \label{eq:Saliency}
    R_i(x) = \frac{\partial f_c(x)}{\partial x_i}
\end{equation}

During BP, when a ReLU activation is encountered, the gradient signals are zeroed out corresponding to the negative values of activations during FP as shown in Equation \ref{eq:Saliency-ReLU}. Thus, we need to store the indices of the negative activation values in order to support BP for a ReLU activation. 

\begin{equation} \label{eq:Saliency-ReLU}
    R_i^L = (f_i^L > 0) \odot R_i^{L+1} \mbox{ , where } R_i^L = \frac{\partial f^{out}}{\partial f_i^L}
\end{equation}

\subsection{DeconvNet}
Deconvolution was originally designed to reconstruct the input of a CNN starting from the network outputs, in an unsupervised manner. It has been widely adopted as an XAI technique owing to its visualization power of most discriminative features. DeconvNet consists of inverse operations of the FP through a CNN. During BP, the convolutional layers are replaced with deconvolutions and max-pooling layers are replaced with unpooling layers. Deconvolution can be viewed as a transposed convolution and hence, DeconvNet boils down to evaluating gradients of the output with respect to the input features. The primary difference compared to vanilla gradients is in the handling of ReLU layer as shown in Equation \ref{eq:DeconvNet-ReLU}. 

\begin{equation} \label{eq:DeconvNet-ReLU}
    R_i^L = (R_i^{L+1} > 0) \odot R_i^{L+1}
\end{equation}

During BP, DeconvNet applies the ReLU function on the gradient values itself. Thus, it does not incur the memory overhead of storing indices of negative activation values during FP. However, by doing so, only those features that have a positive contribution to the model's output are highlighted.

\subsection{Guided Backpropagation}
This method combines the ideas of vanilla backpropagation in Saliency Maps with DeconvNet to make more accurate reconstructions starting from deeper layers of the network. As shown in Equation \ref{eq:GuidedReLU}, Guided Backpropagation zeroes out values corresponding to negative activations (during FP) as well as negative gradients (during BP) when it encounters a ReLU layer. Similar to DeconvNet, the heatmap would only highlight features that positively contribute to the model's output decision. 

\begin{equation}
    \label{eq:GuidedReLU}
    R_i^L = (f_i^L > 0) \odot (R_i^{L+1} > 0) \odot R_i^{L+1}
\end{equation}

In this case, we need to store indices of negative values of the activations during FP to guide the gradient computation across ReLU layers during BP. Thus, the incurred memory overhead is same as that of Saliency Maps.

\section{Hardware Implementation}
Edge devices have limited hardware resources. There is a tight constraint on available memory, bandwidth, compute units as well as the power consumption. This necessitates a tile-based design in order to maximally extract the available parallelism in the algorithm while adhering to the hardware constraints. 

\subsection{Design Overview}
We design a HLS library to support feature attribution for CNNs which typically consist of convolutional layers, pooling, fully connected (FC) layers and non-linear activations such as ReLU. The design is optimized to fit on a small FPGA. The CNN model parameters (weights) as well as the input image are stored in DRAM. Computation is performed in a layer wise manner for both FP and BP phases. Each layer is broken into multiple tiles based on its size. The tiles are loaded into the on-chip buffers from the DRAM using AXI interface. The output tile is stored back into the DRAM once computation is complete. This serves as the input for the next layer of the network. 

\subsection{Convolution Block}
The primary compute kernel in CNNs during both BP and FP is the convolution operation. On-chip buffers are allocated to store the input and filter kernel tiles. The multiply-and-accumulate (MAC) operation is performed on these tiles utilizing the dedicated DSP units on the FPGA. Loop unrolling is performed in order to execute several MACs simultaneously in the same cycles. The loops along the height and width dimension of the input feature maps are unrolled and the corresponding buffers are partitioned accordingly. The unroll factors are configurable at design time. An output stationary dataflow is employed to perform the convolution. The output values are accumulated in-place in the output buffer while iterating over the input tiles. After the output tile computation is complete, it is stored back into DRAM.

\subsection{Vector-Matrix Product Block}
While the convolutional layers in a CNN are used as feature extractors, the FC layers at the end combine those features together to generate the output. FC layers can be mathematically expressed as a vector-matrix multiplication (VMM). To support FC layers, we design a VMM compute block. In order to have a tiled design, on-chip buffers are allocated for the input vector, the weight matrix and the output vector. The input vector as well as the weight matrix are split into tiles and loaded from the DRAM into the corresponding on-chip buffers. Output stationary is employed to accumulate the result in the buffers. The output tile is stored back into DRAM once it is completely evaluated. Partitioning the buffers and unrolling the loop performing the MAC operation enables the design to utilize the available parallelism. 

\subsection{Non-linear Layers}
Apart from convolutional and FC layers, CNNs typically also comprise of other layers such as pooling and non-linear activations. Our HLS library currently supports max-pooling and ReLU activation. The implementation of these layers is designed to support both FP as well as BP.

\textbf{ReLU}. The Rectified Linear Unit (ReLU) activation zeroes out negative values going into the layer. ReLU is implemented via in-place modification of the value in the on-chip output buffers before storing those values back into DRAM. This reduces the data movement between DRAM and on-chip buffers. To support BP, we observe that the gradient of a ReLU activation is a step-function. The gradient value is 1 for positive inputs and 0 for negative. Thus, at a ReLU layer during FP, a 1-bit mask is stored in the on-chip BRAM. This mask has the same dimension as the input feature map to the ReLU layer. This mask is utilized during BP to propagate the local activation gradients.

\begin{figure}[ht!]
\centering
\includegraphics[width=0.8\linewidth]{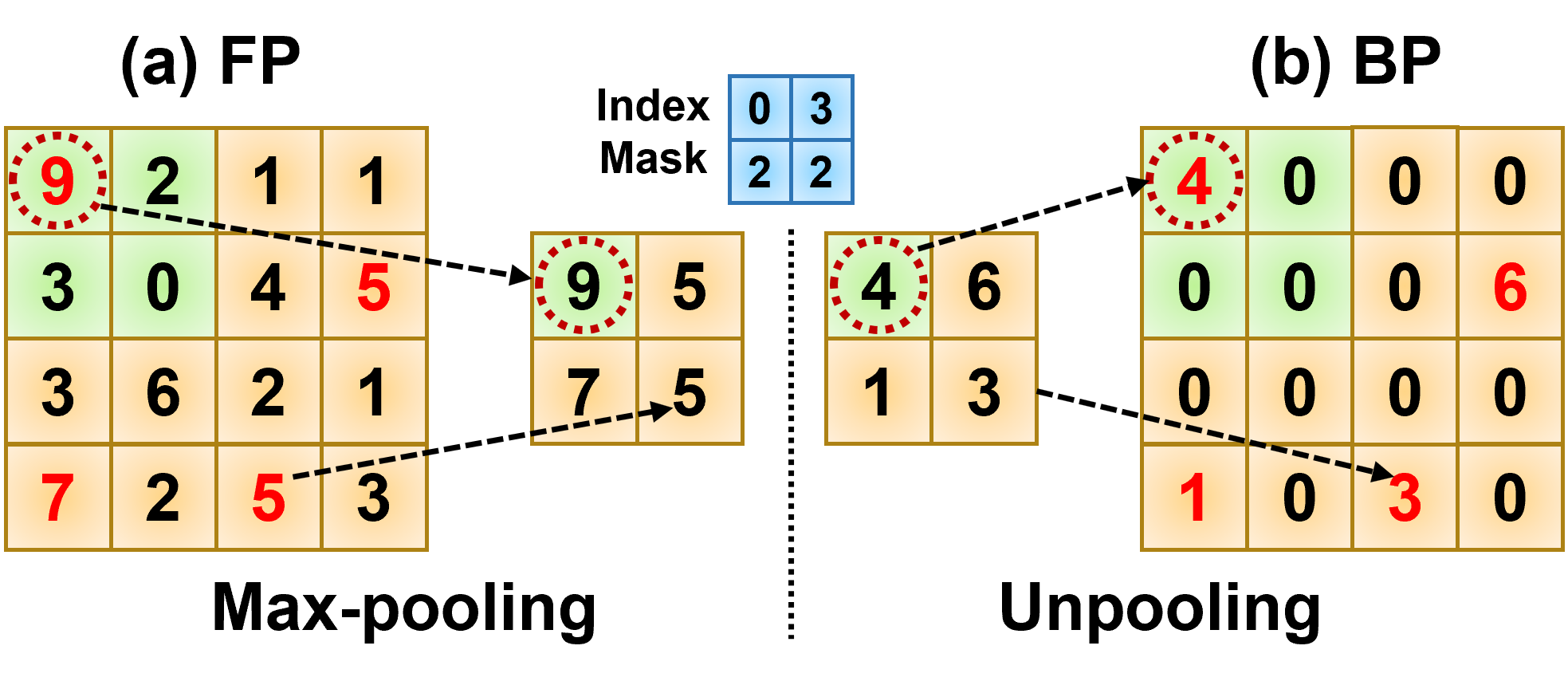}
\caption{Illustration of (a) Maxpooling with 2x2 window and stride of 2 during FP (b) Unpooling of gradient signal at same layer. The 2b index routes the gradient during BP}\label{figure:pooling}
\end{figure}

\textbf{Max-pooling}. Pooling layers reduce the size of feature maps via sub-sampling. A max-pooling layer picks the largest value in the sampling window and passes it to the next layer during FP (Fig. \ref{figure:pooling}a). The window size is typically 2x2 and a stride of 2 ensures no overlap. The implementation of max-pooling is absorbed into the output store operation of the layer that it follows. After the computation of an output tile is complete, the values are scanned based on the pooling window and the stride and only the maximum value is written back into the DRAM. 

\textbf{Unpooling}. Unpooling layers increase the size of feature maps via up-sampling. During BP phase, the gradient across a max-pooling layer is the unpooling operation. The index of the maximum value within the pooling window during FP is stored on-chip. For a 2x2 pooling window, each index is a 2 bit value. The size of the entire index mask is same as the dimension of the output feature map of the max-pooling layer. Unpooling operation uses the cached index value to perform the gradient routing (Fig. \ref{figure:pooling}b) with the remaining values in the pooling window set to 0. 

\subsection{Gradient Computation}
The BP phase for feature attribution requires computation of activation gradients sequentially in a layer by layer manner. The gradient values with respect to input features are evaluated using chain rule for derivatives. Our design is able to reuse the compute blocks designed for FP to perform gradient computation during BP as well. Thus, we have minimal area overhead to support feature attribution in addition to inference. 

\textbf{FC layers}. Since the FP FC layer is a VMM product, its gradient ends up being a matrix-vector product. Thus, in order to re-use the VMM block, the on-chip buffers are loaded in a transpose manner from the DRAM during BP. 

\begin{figure}[ht!]
\centering
\includegraphics[width=\linewidth]{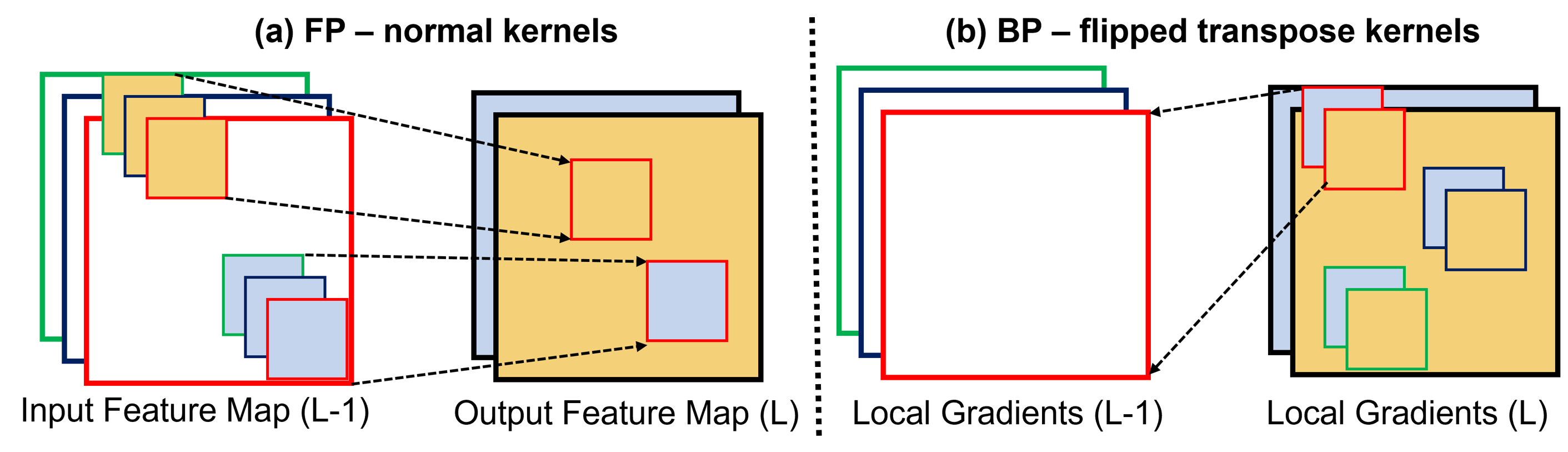}
\caption{(a) Feedforward convolutions (FP) with 3 input channel and 2 output channels with normal kernels (b) Backward convolutions (BP) with 2 input channels and 3 output channels with transposed kernels }\label{figure:conv_bp}
\end{figure}

\textbf{Convolution layers}. The gradient computation with respect to activations for a convolution layers remains a convolution of similar dimension as that during FP. The only difference being (1) the input and output channel dimensions of the layer weight parameters are transposed (Fig. \ref{figure:conv_bp}) and (2) the values of each kernel are flipped by 180$^{\circ}$. We term this as a flipped-transpose convolution during the BP phase. The DRAM access pattern is modified during BP to handle loading the buffers in the required manner. 

\subsection{Scheduling}
The HLS library comprises of template functions that are required to support FP and BP phase for different layers of a CNN. Based on the CNN architecture, appropriate template functions are chosen from the library to execute the model. The layers are scheduled in a sequential manner. The output of each layer is stored into DRAM and is treated as the input for the next layer. The on-chip buffers and compute blocks are fixed at design time. Each layer of the CNN is broken up into tiles based on its size. Starting with the FP phase, the network output is evaluated. Mask values are stored on-chip at non-linear layers. The maximum output value at the last layer is chosen as the output of the network. The BP phase begins at this output value. The DRAM loading functions into the convolution (Table \ref{table:opPhase}) and VMM buffers are chosen according to operation phase. Gradient signals are propagated back to the input features. The feature relevance values and the inference output are evaluated for one input at a time (batch size = 1).

\begin{table}[ht!]
\centering
\caption{Buffer re-use across computational phase}\label{table:opPhase}
\begin{tabular}{|c|c|c|c|}
\hline
\textbf{\begin{tabular}[c]{@{}c@{}}Phase\end{tabular}} & \textbf{Input}                                                               & \textbf{Weight}                                                           & \textbf{Output}                                                            \\ \hline
\textit{\textbf{FP}}                                                & \begin{tabular}[c]{@{}c@{}}Activations\\ (Layer L)\end{tabular}              & \begin{tabular}[c]{@{}c@{}}Normal \\ Kernel\end{tabular}                  & \begin{tabular}[c]{@{}c@{}}Activations\\ (Layer L+1)\end{tabular}          \\ \hline
\textit{\textbf{BP}}                                                & \begin{tabular}[c]{@{}c@{}}Activation \\ Gradient\\ (Layer L+1)\end{tabular} & \begin{tabular}[c]{@{}c@{}}Flipped + \\ Transposed \\ Kernel\end{tabular} & \begin{tabular}[c]{@{}c@{}}Activation \\ Gradient\\ (Layer L)\end{tabular} \\ \hline
\end{tabular}
\end{table}

\subsection{Reconfigurable Design}
The HLS library is designed to be reconfigurable and support different visualization based XAI methods which rely on gradient backpropagation. The memory overhead and the dataflow at the ReLU layers depends on the choice of feature attribution algorithm at design time (Table \ref{table:overhead}). Currently, the library supports three different dataflows: (1) Saliency Map (2) DeconvNet and (3) Guided Backpropagation. Of the three, DeconvNet has the smallest memory overhead while Guided Backpropagation introduces the largest amount of sparsity in intermediate gradient signals.

\begin{table}[ht!]
\centering
\caption{Memory overhead comparison at non-linearities}\label{table:overhead}
\begin{tabular}{|c|c|c|c|}
\hline
\textbf{\begin{tabular}[c]{@{}c@{}}Attribution \\ Method\end{tabular}} & \textbf{\begin{tabular}[c]{@{}c@{}}Saliency\\ Map\end{tabular}} & \textbf{\begin{tabular}[c]{@{}c@{}}Deconv\\ Net\end{tabular}} & \textbf{\begin{tabular}[c]{@{}c@{}}Guided\\ Backpropagation\end{tabular}} \\ \hline
\textit{ReLU Mask}                                                     & Yes                                                             & No                                                            & Yes                                                                       \\ \hline
\textit{Pooling Mask}                                                  & Yes                                                             & Yes                                                           & Yes                                                                       \\ \hline
% \textit{Sparsity}                                                      & Low                                                             & Low                                                           & High                                                                      \\ \hline
\end{tabular}
\end{table}

\section{Results}
\subsection{Experimental Setup}
To evaluate our hardware design, we train a representative CNN for CIFAR-10 dataset similar to \cite{SeoCNNTrainingFPGA} by stacking commonly used layers. The structure of the CNN is shown in Table \ref{table:CNN structure}. The model size (2.26 MB) is comparable to SqueezeNet\cite{iandola2016squeezenet}, a commonly used DNN model for edge applications. We use PyTorch to train the network achieving 88\% accuracy after 20 epochs.    

\begin{table}[ht!]
\centering
\caption{CNN structure}\label{table:CNN structure}
\begin{tabular}{|l|l|l|l|}
\hline
{\color[HTML]{000000} \textbf{Input Shape}} & {\color[HTML]{000000} \textbf{Layer (type)}} & {\color[HTML]{000000} \textbf{Output Shape}} & {\color[HTML]{000000} \textbf{\# parameters}} \\ \hline
{[}3,32,32{]}                             & Conv2d                                       & {[}32,32,32{]}                             & 896                                                  \\ \hline
{[}32,32,32{]}                            & Conv2d                                       & {[}32,32,32{]}                             & 9248                                                 \\ \hline
{[}32,32,32{]}                            & MaxPool2d                                    & {[}32,16,16{]}                             &                                                      \\ \hline
{[}32,16,16{]}                            & Conv2d                                       & {[}64,16,16{]}                             & 18496                                                \\ \hline
{[}64,16,16{]}                            & Conv2d                                       & {[}64,16,16{]}                             & 36,928                                               \\ \hline
{[}64,16,16{]}                            & Maxpool2d                                    & {[}64,8,8{]}                               &                                                      \\ \hline
{[}64,8,8{]}                              & FC                                           & {[}128{]}                                  & 524416                                               \\ \hline
{[}128{]}                                 & ReLU                                         & {[}128{]}                                  &                                                      \\ \hline
{[}128{]}                                 & FC                                           & {[}10{]}                                   & 1290                                                 \\ \hline
\end{tabular}
\end{table}

The FPGA accelerator is designed for this network using the HLS library and synthesized using Xilinx Vitis HLS tool at a target frequency of 100 MHz. The configurable data precision is set to 16-bit fixed point for activations, weights and gradient values. To demonstrate the flexibility of our HLS library, we synthesize our design on three different FPGAs: (1) Pynq-Z2 (2) Ultra96-V2 (3) ZCU104. While (1) is based on Xilinx Zynq-7000 SoC, (2) \& (3) are based on Xilinx Zynq Ultrascale+ MPSoCs. The resource availability and power consumption of these platforms are comparable with edge devices. The hardware configuration of the synthesized design (which determines the resource utilization and latency) are chosen according to the target FPGA platform. 

\begin{table*}[ht!]
\centering
\caption{Evaluation of the hardware design on different target FPGA platforms.} \label{table:results}
\begin{tabular}{|c|c|cc|cccc|c|}
\hline
\multirow{2}{*}{\textbf{FPGA}}                & \multirow{2}{*}{\textbf{\begin{tabular}[c]{@{}c@{}}Operating\\ Phase\end{tabular}}} & \multicolumn{2}{c|}{\textbf{Unroll Factors}}                 & \multicolumn{4}{c|}{\textbf{Resource Utilization}}                                                                                      & \multirow{2}{*}{\textbf{\begin{tabular}[c]{@{}c@{}}Latency\\ (ms)\end{tabular}}} \\ \cline{3-8}
                                              &                                                                                     & \multicolumn{1}{c|}{\textit{N$_{oh}$}}  & \textit{N$_{ow}$}  & \multicolumn{1}{c|}{\textit{BRAM}} & \multicolumn{1}{c|}{\textit{DSP}} & \multicolumn{1}{c|}{\textit{FF}}  & \textit{LUT}               &                                                                                  \\ \hline
\multirow{3}{*}{\textit{\textbf{Pynq-Z2}}}    & FP                                                                                  & \multicolumn{1}{c|}{\multirow{3}{*}{4}} & \multirow{3}{*}{4} & \multicolumn{1}{c|}{10 (3\%)}      & \multicolumn{1}{c|}{32 (14\%)}    & \multicolumn{1}{c|}{18.6K (17\%)} & 38.4K (72\%)               & 43.53                                                                            \\ \cline{2-2} \cline{5-9} 
                                              & FP+BP                                                                               & \multicolumn{1}{c|}{}                   &                    & \multicolumn{1}{c|}{11 (3\%)}      & \multicolumn{1}{c|}{33 (15\%)}    & \multicolumn{1}{c|}{26.7K (25\%)} & 52.9K (99\%)               & 66.75                                                                            \\ \cline{2-2} \cline{5-9} 
                                              & \multicolumn{1}{l|}{Overhead}                                                       & \multicolumn{1}{c|}{}                   &                    & \multicolumn{1}{l|}{1}             & \multicolumn{1}{l|}{1}            & \multicolumn{1}{l|}{8.1K}         & \multicolumn{1}{l|}{14.5K} & \multicolumn{1}{l|}{23.22}                                                       \\ \hline
\multirow{3}{*}{\textit{\textbf{Ultra96-V2}}} & FP                                                                                  & \multicolumn{1}{c|}{\multirow{3}{*}{4}} & \multirow{3}{*}{8} & \multicolumn{1}{c|}{10 (2\%)}      & \multicolumn{1}{c|}{48 (13\%)}    & \multicolumn{1}{c|}{19.2K (13\%)} & 47.8K (67\%)               & 24.56                                                                            \\ \cline{2-2} \cline{5-9} 
                                              & FP+BP                                                                               & \multicolumn{1}{c|}{}                   &                    & \multicolumn{1}{c|}{11 (2\%)}      & \multicolumn{1}{c|}{49 (13\%)}    & \multicolumn{1}{c|}{25.6K (18\%)} & 62.9K (89\%)               & 39.96                                                                            \\ \cline{2-2} \cline{5-9} 
                                              & \multicolumn{1}{l|}{Overhead}                                                       & \multicolumn{1}{c|}{}                   &                    & \multicolumn{1}{l|}{1}             & \multicolumn{1}{l|}{1}            & \multicolumn{1}{l|}{6.4K}         & \multicolumn{1}{l|}{15.1K} & \multicolumn{1}{l|}{15.4}                                                        \\ \hline
\multirow{3}{*}{\textit{\textbf{ZCU104}}}     & FP                                                                                  & \multicolumn{1}{c|}{\multirow{3}{*}{8}} & \multirow{3}{*}{8} & \multicolumn{1}{c|}{10 (1\%)}      & \multicolumn{1}{c|}{96 (5\%)}     & \multicolumn{1}{c|}{27.2K (5\%)}  & 68.1K (29\%)               & 15.32                                                                            \\ \cline{2-2} \cline{5-9} 
                                              & FP+BP                                                                               & \multicolumn{1}{c|}{}                   &                    & \multicolumn{1}{c|}{11 (1\%)}      & \multicolumn{1}{c|}{97 (5\%)}     & \multicolumn{1}{c|}{34.9K (7\%)}  & 85.7K (37\%)               & 26.37                                                                            \\ \cline{2-2} \cline{5-9} 
                                              & \multicolumn{1}{l|}{Overhead}                                                       & \multicolumn{1}{c|}{}                   &                    & \multicolumn{1}{l|}{1}             & \multicolumn{1}{l|}{1}            & \multicolumn{1}{l|}{7.7K}         & \multicolumn{1}{l|}{17.6K} & \multicolumn{1}{l|}{11.05}                                                       \\ \hline
\end{tabular}
\end{table*}

\subsection{Analysis}
\textbf{Design Configuration}. The configurable parameters for the synthesized design are the buffer sizes for the convolution and VMM compute blocks. The input/output buffers of the convolution block are partitioned along the height (width) dimension with a factor N\textsubscript{\textit{oh}} (N\textsubscript{\textit{ow}}). The DSP utilization for the convolution block is N\textsubscript{\textit{oh}} $\times$ N\textsubscript{\textit{ow}} owing to parallel MAC operation. For the VMM block, the buffer size is set to 16/32 based on available resources and the DSP utilization is equal to the same. Table \ref{table:results} shows the configurations chosen for the target FPGA boards. The hardware configuration remains same for both FP and BP phase of feature attribution for CNNs since we efficiently reuse the compute blocks. 

\textbf{Resource Utilization}. Table \ref{table:results} shows the breakdown of the hardware resource utilization for inference (only FP) and feature attribution (FP+BP) for different configurations. We observe that the utilization of BRAM (memory) and DSP (compute) shows negligible change when we add a BP phase to support feature attribution. Thus, our design efficiently reuses the inference hardware to support feature attribution. DSP utilization is in accordance to the design configuration parameters chosen prior to synthesis. The increase in the FF and LUT utilization on adding the BP phase is attributed to the additional logic required for scheduling the layers. Feature attribution would make the scheduler go through the network layers twice compared to just once during inference. High LUT consumption, which is the limiting factor for further speedup in each configuration, is attributed to two reasons (1) partitioning of on-chip buffers for parallel read/write access, and (2) multiplexers for loading the on-chip buffers from different layer parameters in the DRAM. (1) is required to extract speedup from parallelism and (2) is required to efficiently reuse the hardware by changing DRAM access patterns during FP and BP.   

\textbf{Latency} Table \ref{table:results} shows the latency for different hardware configurations obtained via simulation of the synthesized design at a 100 MHz clock. Larger loop unroll factors lead to higher parallelism in the MAC computation. As expected, the latency is lower for larger unroll factors on FPGAs with more available resources. The latency breakdown is provided for running inference (FP) and feature attribution (FP+BP). The overhead of supporting feature attribution in addition to inference manifests in the end-to-end latency of the running the entire network and varies from 50\%-72\% depending on the hardware configuration. On larger FPGAs, the FP and BP phases can be pipelined to improve the throughput of the design by $\approx$1.6$\times$ at the cost of separate compute blocks.

\section{Discussion \& Related Work}
\textbf{Software}. Commonly used DNN software frameworks for CPU/GPU platforms such as Tensorflow and PyTorch implement BP via automatic differentiation. This incurs a large memory overhead since activation values during FP are cached to recursively evaluate gradients during BP using chain rule for derivatives. For the chosen network architecture (Table \ref{table:CNN structure}), the memory overhead is 3.4 Mb. Our design avoid this issue by computing activation gradients in an analytic manner requiring mask bits only at non-linear layers. This reduces memory footprint to 24.7 Kb (137$\times$ lower). This optimization is specific to feature attribution since it does not involve calculating gradients with respect to weight parameters.  

\textbf{Inference Hardware}. Optimized hardware architectures have been proposed to accelerate DNN inference \cite{chen2020survey} and prototyped on FPGA platforms. These designs only support FP phase but XAI techniques require additional computations in the form of gradient backpropagation. Our work highlights how to repurpose inference accelerators to support BP with low resource overhead for visualization based XAI.    
 
\textbf{Training Hardware}. Hardware architectures designed to accelerate DNN training can directly support vanilla gradient based feature attribution \cite{lee2021overview}. However, modifications are required to support other visualization algorithms. These designs are optimized for evaluating weight gradients (WU phase) and incur memory overhead of caching activations during FP. This work highlights that these overheads can be avoided for feature attribution. 
% \cite{parhi} reduces memory access during training by interleaving calculation of weight and activation gradients on a systolic array substrate. However, the optimization is specific to FC layers (multi-layer perceptron) while our design supports end-to-end feature attribution for CNNs.

\textbf{XAI Hardware}. XAI being a relatively nascent field, there is still a dearth of hardware architectures that are specialized for these algorithms. \cite{pan2021hardware} focuses on model distillation based explanation. Model distillation requires training a new interpretable model which mimics the original model's input-output behavior locally. Their target platform is server class TPU/GPU whose energy efficiency make them unsuitable for edge applications. We focus on visualization methods which does not require training new models and is suitable for real-time edge applications. \cite{SeoCNNTrainingFPGA} implements a cyclic weight storage to allow for normal (FP) and transpose (BP) access with no memory overhead. It incurs a logic overhead of address translator block. We observe that this technique is necessary only when all weights are stored in on-chip buffers. Our optimization relies on the data movement that occurs between DRAM and on-chip buffers in tiling based designs. The DRAM access patterns are modified during different compute phases. Thus, our design incurs low logic overheads while also simplifying the hardware design.

\section{Conclusion}
In this paper, we present a HLS based FPGA accelerator for end-to-end XAI for CNNs. The dataflow of multiple feature attribution algorithms is analyzed to design a configurable HLS library supporting three different algorithms. By identifying the difference in the dataflow of inference, feature attribution and training, the gradient computation is optimized to minimize memory overhead. The accelerator is synthesized at 16-bit fixed point precision on multiple FPGAs with varying resource constraints demonstrating flexibility of our HLS library. Analysis of our implementation shows that supporting feature attribution in addition to inference incurs a latency cost while the resource overhead is minimal. Our methodology of reusing allocated memory and compute resources by modifying memory access patterns in inference accelerators and repurpose them to support feature attribution paves the way to enable real-time XAI on edge devices.

% The work presented in this paper proposed a HLS based flexible accelerator for real time feature attribution for edge applications. Our design shown the successful implementation of commonly used layers in CNN with a reconfigurable hardware  which can support both inference and activation gradient backpropagation. We demonstrated the reuse of on-chip compute blocks for both FP and BP phase while having a low memory overhead. The dataflow is flexible enough to support different feature attribution algorithms. Performance comparison with CPU shows 10x improvement in terms of latency.

\bibliographystyle{IEEEtran}
\bibliography{ArXiV_Submission.bbl}

\end{document}